\documentclass[onecolumn,usenatbib]{mn2e}
\usepackage{graphicx}

\begin{document}

\title[Superhumps in AM~CVn Systems]
{Are Superhumps Good Measures of the Mass Ratio for AM CVn Systems?}

\author[Pearson]
{K.\ J.\ Pearson
\\
Department of Physics and Astronomy, Louisiana State University,
202 Nicholson Hall, Baton Rouge, LA 70803-4001, USA. \\
}

\date{Accepted . Received ; 
in original form }

\maketitle

\begin{abstract} 
We extend recent work that included the effect of pressure forces to
derive the precession rate of eccentric accretion
discs in cataclysmic variables to the case of double degenerate systems. We
find that the logical scaling of the pressure force in such systems 
results in predictions of unrealistically high primary masses. Using the 
prototype AM~CVn as a calibrator for the magnitude of the effect, we find 
that there is no scaling that applies consistently to all the systems in the
class. We discuss the reasons for the lack of a superhump period to
mass ratio relationship analogous to that known for SU~UMa systems and 
suggest that this is because these secondaries do not have a single valued 
mass-radius relationship. We highlight
the unreliability of mass-ratios derived by applying the SU~UMa expression to
 the AM~CVn binaries.  
\end{abstract}


\section{Introduction}
\label{sec:intro}

The standard model of a cataclysmic variable (CV) is of a semi-detached
binary consisting of a white dwarf accreting material through
an accretion disc from a late-type
main-sequence secondary. Among the taxonomical classes of CV are the dwarf
novae (DNe) and their sub-class named after the prototype SU~UMa. DNe show 
semi-regular outbursts of increased luminosity (typically 2--5 mag.) lasting  
for a few days and recurring on timescales of tens of days to tens of years.
The SU~UMa systems show occasional superoutbursts with a brighter maximum 
($\sim0.7$~mag.) and longer duration (approximately 5 times) than normal
outbursts. They also show ``superhumps'': a periodicity observed in the 
lightcurve during an outburst occurring on a period ($P_{\rm sh}$) a few
percent longer than the orbital period ($P_{\rm orb}$) \citep{warner95}.

DN outbursts are believed to result from an ionization instability that
sets in when the accretion disc exceeds a critical maximum surface density
($\Sigma_{\rm max}$). On crossing this threshold, the disc switches to a 
``hot'' state with a significantly higher viscosity and mass throughput. This
results in a decrease in the surface density until the disc reaches a second
critical surface density ($\Sigma_{\rm min}$) where the disc can no longer
maintain the hot state. At this point, the disc switches back to a low
viscosity configuration, begins to refill and the cycle repeats
\citep{meyer81,FKR}.

AM~CVn systems are analogues of DNe where the mass-donating secondary is a
helium-rich degenerate or semi-degenerate star. These systems can similarly 
exhibit outbursts with detectable superhumps.

Superhumps arise from an ellipticity induced in the shape of the accretion 
disc by tidal excitation of a resonance with the secondary 
\citep{whitehurst88a}. This elliptical 
pattern has a precession period ($P_{\rm pr}$) much longer than 
$P_{\rm orb}$.
Thus the secondary and the major axis of the disc come into the same 
relative alignment on the beat period between these two which is consequently
just slightly longer than $P_{\rm orb}$. It is the modulation of the
tidal dissipation on this beat period that is picked up observationally 
as $P_{\rm sh}$. 
\citet{lubow91,lubow91b,lubow92} derived the steady state precession rate 
$\omega_{\rm pr}$ for an eccentric disc as: 
\begin{equation}
\omega_{\rm pr}=\omega_{\rm dyn} +\omega_{\rm press}
\end{equation}
where $\omega_{\rm dyn}$ is the dynamical precession frequency and
$\omega_{\rm press}$ is a pressure related term. These authors recognised
the similarity between the precession of the accretion disc and the familiar 
inner Lindblad resonance that produces spiral waves in galactic discs.  
Historically, only the dynamical frequency 
\begin{equation}
\frac{\omega_{\rm dyn}}{\omega_{\rm orb}} =
\frac{q}{\left(1+q\right)^{\frac{1}{2}}}
\left[\frac{1}{2r^{\frac{1}{2}}} \frac{d}{d\!r}
\left( r^{2} \frac{d}{d\!r}\left\{ _{2}F_{1}\left(\frac{1}{2},\frac{1}{2};1;r^2
\right)\right\}\right) \right]
\label{eqn:omdynbasic}
\end{equation}
\citep{hirose90} has been considered important where $q=M_{2}/M_{1}$ is the 
mass ratio, $r$ is the distance from the primary (expressed here as a fraction 
of the separation $d$) and $_{2}F_{1}$ is the hypergeometric function. 
Recently the necessity of including
\begin{equation}
\omega_{\rm press}=-\frac{c^{2} \cot^{2} i}{2 r^{2} \omega_{\rm p}} 
\label{eqn:ompress}
\end{equation}
has been reasserted \citep{murray00}, where $\omega_{\rm p}$ is the orbital 
angular frequency of disc material at a radius $r$ (hereafter measured in
dimensional units) and $i$ is the pitch
angle of the induced spiral wave. This was studied by 
\citet{goodchild06}
with a detailed integration scheme and \citet{pearson06} with an algebraic
approach to derive an improved relationship between $\omega_{\rm pr}$ and the
system parameters. In particular, the latter paper used the observed
values of $P_{\rm sh}$ to derive values for the component masses of
88 CVs. These papers provided a theoretical foundation for the observed
correlation between the observed period excess 
($\epsilon=(P_{\rm sh} -P_{\rm orb})/P_{\rm orb}$) and $q$ \citep{patterson05}.
While the nature of the superhumps in ordinary CVs is well-established,
the application of the theory to AM~CVn systems relies on a ``by analogy''
argument. Here we examine the degree to which this same approach can be 
successfully applied to AM~CVn systems.

\section{Precession Frequency}
\label{sec:theback}

\subsection{Dynamical Precession}

The dynamical precession rate can be evaluated by idealising the response of 
the disc as that of a ring of material at a radius 
$r=j^{-\frac{2}{3}} (1+q)^{-\frac{1}{3}} d$ that has an orbital frequency 
about the primary in a resonance with the secondary. 
The expression given in (\ref{eqn:omdynbasic}) can be then be
reduced to
\begin{equation}
\frac{\omega_{\rm dyn}}{\omega_{\rm orb}}  =
\frac{3}{4j}\frac{q}{1+q}
\sum_{n=1}^{\infty}
\frac{a_{n}}{\left[j^{2}{(1+q)}\right]^{\frac{2(n-1)}{3}}}
\label{eqn:omdynred}
\end{equation}
where
\begin{equation}
a_{n}=\frac{2}{3}(2n)(2n+1)\prod_{m=1}^{n} \left(\frac{2m-1}{2m}\right)^{2}
\label{eqn:coeffrel}
\end{equation}
\citep{pearson03} and $j=3$ is the appropriate resonance. 
If the dynamical term were the only contribution to the precession
rate we could rewrite equation~(\ref{eqn:omdynred}) as
\begin{equation}
P_{\rm pr} = 2.57 \frac{1+q}{q} P_{\rm orb}
\label{eqn:pshofq}
\end{equation} 
by evaluating the summation with a ``typical'' $q=0.16$ 
(cp. \cite{warner95}). The coefficient in (\ref{eqn:pshofq}) differs from the
often used value of $\approx3.85$. As noted by \citet{murray00}, this latter 
value results from an erroneous factor of 2/3 being introduced. This
comes from considering the time for a test-particle to return to the
same relative alignment with the secondary rather than a ring of material
along the resonant orbit. 

The above prescription for the radius assumes that
$\omega_{\rm pr} \ll \omega_{\rm orb}$ in order to reduce the
resonance condition 
\begin{equation}
(j-1)(\omega_{\rm p} -\omega_{\rm pr})=j(\omega_{\rm p} -\omega_{\rm orb})
\end{equation}
to the approximate 
$\omega_{\rm p}\approx j \omega_{\rm orb}$. As such, the infinite sum in
(\ref{eqn:omdynred}) does not carry the accuracy that might normally
be assumed for such a summation. It also tacitly assumes that
the resonance can be represented by the response at a single radius rather
than spread throughout the disc structure.
In any case, we also need to account for the additional pressure related term.

\subsection{General Form of $\omega_{\rm press}$}

A fuller explanation of the following derivation is given in \citet{pearson06} 
but we take this opportunity to correct the numerical value of constants 
given there and to write the equations in a form applicable to
both hydrogen- and helium-dominated discs.

For a disc opacity law
\begin{equation}
\kappa=\kappa_{0} \rho T^{-3.5}
\end{equation}
where
\begin{equation}
\kappa_{0}=2.8\times10^{20}~\mbox{m}^{5}~\mbox{kg}^{-2}~\mbox{K}^{-3.5}
\end{equation}
is appropriate \citep{cannizzo92b}, equation~A3 of \cite{cannizzo92a} can 
be written as
\begin{equation}
\Sigma = \Sigma_{0} \mu_{\rm h} 
\left(\frac{\alpha_{\rm h}}{0.1}\right)^{-\frac{4}{5}} M_{1}^{\frac{1}{4}}
\left(\frac{r}{10^{8}~\mbox{m}}\right)^{-\frac{3}{4}}
\left(\frac{\dot{M}}{10^{-10} M_{\odot} \mbox{y}^{-1}}\right)^{\frac{7}{10}}
\label{eqn:sigmahot}
\end{equation}
where $\Sigma_{0,{\rm H}}=406~\mbox{kg}~\mbox{m}^{-2}$, $M_{1}$ is the 
primary mass in solar units and $\dot{M}$ the mass transport rate
through the disc. The subscripts 
$_{\rm h}$ ($_{\rm c}$) will be used to refer to the disc in its hot (cold) 
state and 
the subscripts $_{\rm H}$ and $_{\rm He}$ will be used for hydrogen- and 
helium-dominated discs respectively. Assuming free-free opacity is dominant,
the relationship
\begin{equation}
\kappa_{0,{\rm ff}}\propto (1+X)(1-Z)
\end{equation}
\citep{bowers84} allows us to scale to 
$\Sigma_{0,{\rm He}}=428~\mbox{kg}~\mbox{m}^{-2}$
by using $X_{\rm H}=0.7$, $Y_{\rm H}=0.27$, $Z_{\rm H}=0.03$, 
$X_{\rm He}=0.0$, $Y_{\rm He}=0.97$, $Z_{\rm He}=0.03$ \citep{tsugawa97}.

\citet{cannizzo88} give the maximum possible surface density in the cold 
state as
\begin{equation}
\Sigma_{\rm max,H}=114~\mbox{kg}~\mbox{m}^{-2} 
\left(\frac{r}{10^{8}~\mbox{m}}\right)^{1.05} M_{1}^{-0.35} 
\alpha_{\rm c}^{-0.86}
\end{equation}
for a H-dominated disc and similarly \citet{tsugawa97}
\begin{equation}
\Sigma_{\rm max,He}=2510~\mbox{kg}~\mbox{m}^{-2} 
\left(\frac{r}{10^{8}~\mbox{m}}\right) M_{1}^{-0.33} 
\left(\frac{\alpha_{\rm c}}{0.3}\right)^{-0.7}
\end{equation}
for one that is He-dominated. For consistency in the form of the
equations, we scale the value of this latter relationship to $\alpha_{\rm c}=1$
and adopt here,
\begin{equation}
\Sigma_{\rm max}=\Sigma_{\rm crit} 
\left(\frac{r}{10^{8}~\mbox{m}}\right)^{1.05}
M_{1}^{-0.35} \alpha_{\rm c}^{-0.86}
\label{eqn:sigmamax}
\end{equation}
where $\Sigma_{\rm crit,H}=114~\mbox{kg}~\mbox{m}^{-2}$ and 
$\Sigma_{\rm crit,He}=1080~\mbox{kg}~\mbox{m}^{-2}$.

Integrating (\ref{eqn:sigmahot}) with respect to radius gives the disc mass 
$M_{\rm d}$
and similarly (\ref{eqn:sigmamax}) gives the maximum disc mass in the
cold state $M_{\rm max}$. If the disc fills to a constant fraction $f$ of its
maximum possible mass, the equality $M_{\rm d}=fM_{\rm max}$ results
in
\begin{eqnarray}
\dot{M} & = & \dot{M}_{0} \left(\frac{f}{0.4}\right)^{1.43}
M_{1}^{0.86} r_{\rm d}^{2.57} \left(\frac{\alpha_{\rm c}}{0.02}\right)^{-1.23}
\left(\frac{\alpha_{\rm h}}{0.1}\right)^{1.14} \mu_{\rm h}^{-1.07}
\end{eqnarray}
where $\dot{M}_{0,{\rm H}}=
2.62\times10^{-8}~\mbox{kg}~\mbox{s}^{-1}~\mbox{m}^{-2.57}$ and
$\dot{M}_{0,{\rm He}}=
60.3\times10^{-8}~\mbox{kg}~\mbox{s}^{-1}~\mbox{m}^{-2.57}$. 
Substituting this into equation~A1 of \cite{cannizzo92a} gives
\begin{equation}
T=T_{0} 
\mu_{\rm h}^{-0.071} \left(\frac{\alpha_{h}}{0.1}\right)^{0.142}
\left(\frac{\alpha_{\rm c}}{0.02}\right)^{-0.369}
\left(\frac{f}{0.4}\right) r_{\rm d}^{0.771} M_{1}^{-0.258} 
\omega_{\rm p}^{\frac{1}{2}}
\end{equation}
where $T_{0,\rm H}=0.246~\mbox{K}~\mbox{m}^{-0.771}~\mbox{s}^{\frac{1}{2}}$ 
and $T_{0,{\rm He}}=0.597~\mbox{K}~\mbox{m}^{-0.771}~\mbox{s}^{\frac{1}{2}}$.
Hence,
\begin{eqnarray}
c^{2} & = & \frac{\gamma k T}{\mu_{\rm h} m_{\rm H}} \\
      & = & c_{0}^{2} 
\mu_{\rm h}^{-1.071} \left(\frac{\alpha_{h}}{0.1}\right)^{0.142}
\left(\frac{\alpha_{\rm c}}{0.02}\right)^{-0.369}
\left(\frac{f}{0.4}\right) r_{\rm d}^{0.771} M_{1}^{-0.258} 
\omega_{\rm p}^{\frac{1}{2}} \label{eqn:soundspeed}
\end{eqnarray}
where $c_{0,{\rm H}}^{2}= 
3.40\times10^3~\mbox{m}^{1.229}~\mbox{s}^{-\frac{3}{2}}$ and
$c_{0,{\rm He}}^{2}= 
8.27\times10^3~\mbox{m}^{1.229}~\mbox{s}^{-\frac{3}{2}}$.
Putting this into (\ref{eqn:ompress}), we can further
make further eliminations:
\begin{equation} 
r_{\rm d}=\beta R_{{\rm L},1}=\beta E(q^{-1}) d
\end{equation}
where
\begin{equation}
E(q^{-1})\equiv\frac{0.49 q^{-\frac{2}{3}}}{0.6 q^{-\frac{2}{3}}
+\ln(1+q^{-\frac{1}{3}})}
\end{equation}
\citep{eggleton83}.
The resonance requires
\begin{equation}
\omega_{\rm p}= 3 \omega_{\rm orb}=
3 \left\{GM_{1}\left(1+q\right)M_{\odot}\right\}^{\frac{1}{2}}
d^{-\frac{3}{2}}
\label{eqn:rescond}
\end{equation}
and
\begin{equation}
r=3^{-\frac{2}{3}} \left(1+q\right)^{-\frac{1}{2}}d.
\label{eqn:resrad}
\end{equation}
Equating the radius of the secondary $R_{2}$ to the size of its
Roche lobe $R_{{\rm L},2}=E(q)d$, we can also substitute for the separation
\begin{equation}
d = \frac{R_{2}}{E(q)}.
\label{eqn:dofrq}
\end{equation}
The final expression for the pressure term is thus
\begin{equation}
\frac{\omega_{\rm press}}{\omega_{\rm orb}} = 
-3^{\frac{5}{6}}
\eta_{0}
\cot^{2} i 
\,\mu_{\rm h}^{-1.071}
\left(\frac{\alpha_{\rm h}}{0.1}\right)^{0.142}
\left(\frac{\alpha_{\rm c}}{0.02}\right)^{-0.369}
\left(\frac{f}{0.4}\right)^{0.429}
\left(\frac{\beta}{0.9}\right)^{0.771}
M_{1}^{1.008}
(1+q)^{-\frac{1}{12}} 
\frac{E(q^{-1})^{0.771}}{E(q)^{1.021}} R_{2}^{1.021}
\label{eqn:genpressterm}
\end{equation} 
where $\eta_{0,{\rm H}}=1.27\times10^{-12}~\mbox{m}^{-1.021}$ and 
$\eta_{0,{\rm He}}=3.08\times10^{-12}~\mbox{m}^{-1.021}$.
To make further progress we need a form for the mass-radius relation 
$R_{2}(M_{2})$.

\subsubsection{Main Sequence Secondary}

Using the main sequence mass-radius relationship
\begin{equation}
R_{2}=0.91M_{2}^{0.75} R_{\odot}
\label{eqn:msmrrel}
\end{equation}
\citep{smith98}, (\ref{eqn:genpressterm}) can be written as
\begin{equation}
\frac{\omega_{\rm press}}{\omega_{\rm orb}} = -3^{\frac{5}{6}}
\frac{E(q^{-1})^{0.771}}{E(q)^{1.021}}
\frac{q^{0.766}}{(1+q)^{\frac{1}{12}}} \eta_{\rm H}
\label{eqn:eqnpressfinal}
\end{equation}
where
\begin{equation}
\eta_{\rm H}=1.23\times10^{-3} \cot^{2}i
\left(\frac{\alpha_{\rm h}}{0.1}\right)^{0.142}
\left(\frac{\alpha_{\rm c}}{0.02}\right)^{-0.369}
\mu_{\rm h}^{-1.071}
\left(\frac{f}{0.4}\right)^{0.429}
\left(\frac{\beta}{0.9}\right)^{0.771}
M_{1}^{-0.242}.
\label{eqn:etaH}
\end{equation}
Assuming $\eta_{\rm H}$ to be a constant ($\overline{\eta}_{\rm H}$),
\citet{pearson06} found an excellent best fit value of 
$\overline{\eta}_{\rm H}=0.0109$ for those superhumping CVs with independently 
determined values of $q$.
The summation of (\ref{eqn:omdynred}) and (\ref{eqn:genpressterm}) results in 
an algebraically unwieldy, but numerically easy to invert, expression for 
$\frac{\omega_{\rm pr}}{\omega_{\rm orb}}(q)$.

While a constant $\overline{\eta}_{\rm H}$ produced an excellent fit to the 
data, an accurate calculation of the effect in individual systems should 
include the $M_{1}$ dependence of 
$\eta_{\rm H}$ ie.
\begin{equation}
\eta_{\rm H}=
\left(\frac{M_{1}}{\overline{M}_{1}}\right)^{-0.242}\overline{\eta}_{\rm H}
\end{equation}
where $\overline{M}_{1}=0.76$ is the weighted mean primary mass of the sample 
used to determine the value of $\overline{\eta}_{\rm H}$.
This implies that the coefficient of
proportionality between $P_{\rm sh}$ and $P_{\rm orb}$ is no longer
strictly a function of $q$ alone 
(cf. equation (\ref{eqn:pshofq})).  However, 
the combination of (\ref{eqn:rescond}), (\ref{eqn:dofrq}) and 
(\ref{eqn:msmrrel}) produces a second 
constraint $P_{\rm orb}=P_{\rm orb}(M_{2},q)$. We can thus solve simultaneously
the  $P_{\rm orb}(M_{2},q)$ and $P_{\rm sh}(M_{1},M_{2})$
expressions for both $M_{1}$ and $M_{2}$ together. In principle, this will 
change the 
value of $\overline{M}_{1}$ and the expressions should be iterated upon for
a fully self-consistent solution. However, for a main-sequence secondary,
allowing for this dependence produces masses that differ by only a few
percent from those in Table~4 of \citet{pearson06}. The equivalent 
correction for systems with a (semi-)degenerate secondary is more significant.

\subsubsection{Degenerate and Semi-degenerate Secondary}
\label{sec:degsec}

The approach of the previous section can be applied 
{\it mutatis mutandis} to a secondary with a non-main-sequence structure.
Equation (\ref{eqn:eqnpressfinal}) remains applicable but now
\begin{eqnarray}
\eta_{\rm He}& =& 2.67 \left(\frac{\mu_{\rm h,He}}{\mu_{h,H}}\right)^{-1.071}
\left(\frac{R_{2}}{R_{\odot}}\right)^{1.021} M_2^{-0.766} \eta_{\rm H}\\
 &=& 0.0127 \left(\frac{R_{2}}{R_{\odot}}\right)^{1.021} M_{2}^{-0.766}
\left(\frac{M_{1}}{\overline{M}_{1}}\right)^{-0.242}
\label{eqn:etaHe}
\end{eqnarray}
using the compositions given earlier to derive $\mu_{\rm h,H}=0.618$ and
$\mu_{\rm h,He}=1.347$ and the previously found best value of
$\overline{\eta}_{\rm H}$.

Two forms of the mass-radius relation have been widely used in the literature
corresponding to two possible formation channels for these systems 
\citep{nelemans01a}. If the donor is the remnant core of a low-mass helium star
that begins mass-transfer before helium burning, a zero temperature degenerate 
structure is appropriate.
For a fully degenerate helium white dwarf secondary the mass-radius relation is
\begin{equation}
\frac{R_{2}}{R_{\odot}}=0.0106-0.0064 \ln M_{2} 
+ 0.0015 M_{2}^{2}
\label{eqn:ZSeqn}
\end{equation}
\citep{zapolsky69,rappaport84}. Alternatively, the donor star may
begin helium burning before the onset of mass-transfer. 
In this case (a ``semi-degenerate'' secondary), the 
mass-radius relationship has been approximated by
\begin{equation}
\frac{R_{2}}{R_{\odot}}=b M_{2}^{-\alpha}
\label{eqn:SDeqn}
\end{equation}
where \citet{tutukov89} have $b=0.043,\alpha=0.062$ (hereafter TF parameters) 
and \citet{savonije86} found $b=0.029,\alpha=0.19$ (hereafter SKH parameters).

There also exists a form due to Eggleton that attempts to reconcile a low-mass
appropriate form in (\ref{eqn:ZSeqn}) with that due to \citet{nauenberg72}
for higher masses
\begin{eqnarray}
\frac{R_{2}}{R_{\odot}} & = &
0.0114\left[\left(\frac{M_{2}}{M_{\rm Ch}}\right)^{-\frac{2}{3}}
-\left(\frac{M_{2}}{M_{\rm Ch}}\right)^{\frac{2}{3}}\right]^{\frac{1}{2}}
\left[1+3.5\left(\frac{M_{2}}{M_{p}}\right)^{-\frac{2}{3}}
+\left(\frac{M_{2}}{M_{p}}\right)^{-1}\right]^{-\frac{2}{3}}
\end{eqnarray}
where $M_{\rm Ch}=1.44$ is the Chandrasekhar mass and the constant
$M_{p}=0.00057$ \citep{verbunt88}.

\section{Application to AM~CVn systems}

The results of applying the method outlined in section \ref{sec:degsec} 
to the AM~CVn systems are shown in Table~1. Also
given is the ``dynamical only'' result
(ie. $\omega_{\rm press}=0$) for comparison. The resulting mass ratio
for AM~CVn is lower in every case than the normally quoted value $q=0.087$ (eg.
\citet{nelemans01a}),
as we have used the corrected, full expression for the dynamical expression
(equation \ref{eqn:omdynred}). The effect of the addition of the pressure term 
is to require a larger mass-ratio to achieve the same precession rate. 
Since $M_{2}$ is largely determined by $P_{\rm orb}$,
this in turn requires a smaller value for $M_{1}$. This effect notwithstanding,
we see that the semi-degenerate TF model produces unacceptably high $M_{1}$
values. In contrast, the degenerate relationship produces uncomfortably low 
values.

\begin{table*}
\label{tab:amvan}
\begin{minipage}{170mm}
\begin{center}
\caption{Component masses derived from the application of the relationship
established for SU~UMa systems to all the AM~CVn systems with  measured 
$P_{\rm orb}$ and $P_{\rm sh}$ listed in \protect\citet{rkcat}.}
\begin{tabular}{lccccccccccccll}\hline
System      & \multicolumn{3}{c}{TF (dyn. only)} 
	    & \multicolumn{3}{c}{TF } 
            & \multicolumn{3}{c}{Degenerate } 
            & \multicolumn{3}{c}{Eggleton } 
	    & \multicolumn{1}{c}{$P_{\rm orb}$} 
            & \multicolumn{1}{c}{$\epsilon$} \\
	    & $q$ & $M_{1}$ & $M_{2}$ 
	    & $q$ & $M_{1}$ & $M_{2}$ 
            & $q$ & $M_{1}$ & $M_{2}$ & $q$ & $M_{1}$ & $M_{2}$ 
	    & \multicolumn{1}{c}{(s)}    & \\\hline
AM CVn      & 0.056 & 2.02  & 0.113 & 0.064 & 1.76  & 0.114 
            & 0.078 & 0.42  & 0.033 & 0.076 & 0.48  & 0.036 
            & 1028.8 \protect\footnotemark[1] \protect\footnotemark[2]
	    & 0.0218 \protect\footnotemark[1] \\
HP Lib      & 0.036 & 2.72  & 0.099 & 0.044 & 2.28  & 0.100 
	    & 0.056 & 0.54  & 0.030 & 0.054 & 0.62  & 0.033 
	    & 1102.7 \protect\footnotemark[3] 
            & 0.0144 \protect\footnotemark[3] \protect\footnotemark[4] \\
CR Boo      & 0.027 & 2.22  & 0.060 & 0.038 & 1.59   & 0.061 
	    & 0.058 & 0.36  & 0.021 & 0.053 & 0.45  & 0.024 
            & 1471.3 \protect\footnotemark [5]
            & 0.0109 \protect\footnotemark [6]\\
KL Dra      & 0.049 & 1.20  & 0.059 & 0.066 & 0.91  & 0.060 
	    & 0.094 & 0.22  & 0.020 & 0.089 & 0.27  & 0.024 
            & 1501.8 \protect\footnotemark[7]
            & 0.0193 \protect\footnotemark[7] \\
V803 Cen    & 0.009 & 5.32  & 0.050 & 0.017 & 2.93  & 0.051 
	    & 0.034 & 0.54  & 0.018 & 0.030 & 0.71  & 0.021 
            & 1612.0 \protect\footnotemark[8] 
            & 0.00381 \protect\footnotemark[8] \protect\footnotemark[9] \\
CP Eri      & 0.021 & 2.17  & 0.047 & 0.035 & 1.35  & 0.048 
	    & 0.060 & 0.28  & 0.017 & 0.054 & 0.37  & 0.020 
            & 1701.2 \protect\footnotemark[8]
            & 0.00863 \protect\footnotemark[8] \\
V406 Hya    & 0.017 & 2.06  & 0.034 & 0.036 & 0.99  & 0.035 
            & 0.076 & 0.18  & 0.014 & 0.063 & 0.26  & 0.016 
            & 2027.8 \protect\footnotemark[10]
            & 0.00673 \protect\footnotemark[11] \\\hline
\end{tabular}
\end{center}
\protect\footnotetext[1]{\protect\citet{skillman99}}
\protect\footnotetext[2]{\protect\citet{nelemans01}}
\protect\footnotetext[3]{\protect\citet{patterson02}}
\protect\footnotetext[4]{\protect\citet{odonoghue94}}
\protect\footnotetext[5]{\protect\citet{provencal97}}
\protect\footnotetext[6]{\protect\citet{patterson97}}
\protect\footnotetext[7]{\protect\citet{wood02}}
\protect\footnotetext[8]{\protect\citet{patterson01}}
\protect\footnotetext[9]{\protect\citet{kato04}}
\protect\footnotetext[10]{\protect\citet{roelofs06}}
\protect\footnotetext[11]{\protect\citet{woudt03}}
\end{minipage}
\end{table*}

Recent spectroscopic observations of the {HeI}~4471 line in AM~CVn 
have yielded the 
first direct measurement of $q$ for an AM~CVn system \citep{roelofs06}. 
Disappointingly, their value of $q=0.18$ is higher than
any of those predicted by this superhump analysis. It is
possible that the true effect of the pressure term for AM~CVn is larger than
that predicted from the above derivation but this begs the question
of which unaccounted for factor has changed between the SU~UMa and AM~CVn
systems? To achieve a result of $q=0.18$ for the TF mass-radius relationship 
would 
require $\eta_{\rm He}$ a factor $\sim7$ larger
than expected by the extension of the SU~UMa expression. Examining
the terms in equations (\ref{eqn:etaH}) and (\ref{eqn:etaHe}) it is
difficult to see any dependencies, even conspiring together, that could produce
such a factor except $\cot^{2} i$. The best fit value of 
$\overline{\eta}_{\rm H}$ for SU~UMa systems would correspond to a pitch angle 
$i=24^{\circ}$ with the given fiducial parameters. This is 
sufficiently large that it may violate the tight-winding approximation that
was used to derive equation~(\ref{eqn:ompress}). A systematic change in the
value of $i$ to $10^{\circ}$ would produce the required factor.
However, while this would give the ``correct''
result for AM~CVn, a convergent solution that satisfies both the $P_{\rm orb}$
and $P_{\rm sh}$ constraints simultaneously cannot then be found for 
any other system except HP~Lib. The derived values
for these two systems are listed in Table~2 calculated
with the required scaling of $\omega_{\rm pr}$ for each mass-radius 
relationship to achieve an AM~CVn mass ratio of $q=0.18$.

\begin{table*}
\label{tab:amall}
\begin{minipage}{160mm}
\begin{center}
\caption{Derived component masses after scaling the pitch angle $i$ in the 
pressure term to give an AM~CVn mass ratio $q=0.18$ for each mass-radius 
relation.}
\begin{tabular}{lcccccccccccc}\hline
System      & \multicolumn{3}{c}{TF ($i=10^{\circ}$)} 
            & \multicolumn{3}{c}{SKH ($i=11^{\circ}$)} 
            & \multicolumn{3}{c}{Degenerate ($i=15^{\circ}$)} 
            & \multicolumn{3}{c}{Eggleton ($i=15^{\circ}$)} \\
	    & $q$ & $M_{1}$ & $M_{2}$ & $q$ & $M_{1}$ & $M_{2}$ 
            & $q$ & $M_{1}$ & $M_{2}$ & $q$ & $M_{1}$ & $M_{2}$ \\\hline
AM CVn      & 0.180 & 0.649  & 0.117 & 0.180 & 0.517  & 0.093 
	    & 0.180 & 0.186  & 0.033 & 0.180 & 0.206  & 0.037 \\
HP Lib      & 0.164 & 0.633  & 0.104 & 0.158 & 0.539  & 0.085 
	    & 0.158 & 0.194  & 0.031 & 0.156 & 0.218  & 0.034 \\
\hline
\end{tabular}
\end{center}
\end{minipage}
\end{table*} 

While simulations of low $q$ binaries \citep{truss07} do show a correlation 
between $i$ and $q$, the trend is
for $i$ to increase as $q$ decreases, ie. the exact opposite of what would be 
required to explain the change from SU~UMa to AM~CVn systems. Further,
the same simulations show that the structure of accretion discs is largely
unaffected by a significant decrease in the mass transfer rate that might
be expected for systems with low-mass secondaries ruling out this as a possible
cause. The simulations also show a change in 
the orientation of the accretion disc with respect to the line of centres
of the two stars. This rotation angle $\theta_{\rm rot}$ appears to decrease 
with decreasing $q$ for $q<0.1$. The changes in $i$ and $\theta_{\rm rot}$
may combine to explain why the discs in these simulations also show a tendency 
to be less centrally condensed at low $q$. In such a situation, the 
use of the \citet{cannizzo92a} profiles above may not be valid for such 
extreme systems. The effect appears to be sufficiently small however 
that it is not a viable explanation for the systematic change between the two 
classes and doubtful whether it is the explanation of the differences 
within the AM~CVn group.

A further factor that might be  considered is whether the larger size of the 
primary's Roche lobe in extreme mass ratio systems might allow the
disc to grow sufficiently that it could access the  $j=2$ resonance. 
Leaving aside the unique character of the $j=3$ resonance that enable
it to be excited \citep{lubow91}, evaluating $\omega_{\rm dyn}$ with $j=2$
would actually cause the inferred $q$ to be even smaller 
(cf. \citet{pearson03} that invoked resonances of higher $j$ to explain a 
larger than expected mass ratio for AM CVn). Similarly, our 
analysis has characterised the disc response by the properties at a single 
radius. In reality, the precession is a collective property of the whole
disc but perhaps we can find an effective radius $r_{\rm eff}$ (not 
necessarily at a resonance) that gives good results? In summary, it is possible
to achieve $q=0.18$ for AM~CVn with $r_{\rm eff}=r_{4}$ and a degenerate
mass-radius relationship. However, the value of $M_{1}=0.195$ would then be 
unacceptably  small. In contrast, $r_{\rm eff}=r_{4}$ and the TF
semi-degenerate secondary produces an acceptable $M_{1}=0.984$ but
a too small $q=0.118$. We can satisfy both $M_{1}$ and $q$ simultaneously with
$r_{\rm eff}=r_{5}$ and a TF semi-degenerate secondary that results in
$q=0.19$ and $M_{1}=0.607$. However, applying this to other systems then 
produces unrealistically small values for $M_{1}$. These examples, though not 
exhaustive, strongly suggest that it is not possible to find a suitable 
$r_{\rm eff}$, $M(R)$ combination. A study along the lines of 
\citet{goodchild06} integrating across the whole disc structure would prove 
useful.

The structure models are based on axisymmetric, vertically averaged but 
quite general accretion disc equations with mean molecular weight and opacity 
as parameters \citep{cannizzo92a,shakura73}. The critical
transition temperatures are probably less well understood, although even here 
the arguments are sufficiently general that it is difficult to see that they
could be so wrong as to explain the required change. We
have tested the possibility that the proportionality coefficients were
in error by trying to scale the results to give the correct result for AM CVn.
Since this is not possible, the functional forms of the final
equations themselves must be significantly in error. 

Probably the most crucial issue that should be considered is whether 
following the precedent of using a single mass-radius relation,
such as those in equations (\ref{eqn:ZSeqn}) or (\ref{eqn:SDeqn}),
is appropriate for all the AM~CVn secondaries.
Figure 1 of \citet{deloye05} shows that detailed models of the structure 
of the secondary produce a wide variation
in radius for a given mass depending on the object's temperature. 
At the $P_{\rm orb}$ of AM~CVn, the majority of tracks are
constrained to the region close to the fully degenerate relationship, 
but they increasingly diverge for
smaller values of $M_{2}$. The derived values of $M_{1}$ in Table~2, however,
seem more reasonable for the semi-degenerate forms of $M_{2}(R_{2})$. In this 
case, the aforementioned figure would suggest a surprisingly hot secondary 
($\log T_{2}\sim7.0$--$7.5$).

Regardless of the exact cause, the inability to find a universal coefficient 
for the strength of the pressure
term that can be applied to all AM~CVn systems forces us to conclude that 
there is
no general relationship for these systems analogous to that found for the 
SU~UMa binaries. The most likely reason is the lack of a single valued 
relationship between the secondary's mass and radius that makes it impossible
to convert an expression involving $P_{\rm orb}$ into one for $M_{2}$.

\section{Conclusion}

We have shown how the extension of the theoretical relation relating 
$P_{\rm sh}$ to $q$ that works well for ordinary CVs appears to break down 
badly
when applied to AM~CVn systems. Hence, the values derived even by extrapolating
the empirical relation from SU~UMa to AM~CVn binaries should be treated
with extreme caution. The breakdown seems most likely to arise from
the wide range of radii a secondary of a given mass can adopt. This may be 
exacerbated by a change in the accretion disc structure at very low mass 
ratios.
Enhanced numerical studies that would enable such accretion disc profiles to
be determined in terms of the system parameters may allow us to 
assess the importance of this contribution. Deriving a temperature 
for such faint objects against the glare of the
accretion disc would be very difficult, although in principle, it
would enable us to place the secondary on an appropriate 
theoretical $M_{2}(R_{2})$ track

\section*{ACKNOWLEDGEMENTS}

I thank Juhan Frank for illuminating  and stimulating discussions on the 
nature of the superhump resonance.

\end{document}